\documentclass[10pt,conference]{IEEEtran}
\usepackage[utf8]{inputenc}
\usepackage[T1]{fontenc}
\usepackage{lmodern}
\usepackage{amsmath,amssymb}
\usepackage{booktabs}
\usepackage{array}
\usepackage{multirow}
\usepackage{url}
\usepackage[hidelinks]{hyperref}
\usepackage{graphicx}
\usepackage{xcolor}
\usepackage{tikz}
\usetikzlibrary{positioning,arrows.meta,shapes.geometric,fit,backgrounds}
\usepackage{enumitem}
\usepackage{balance}

\title{From Human Interfaces to Agent Interfaces:\\
Rethinking Software Design in the Age of AI-Native Systems}

\author{
\IEEEauthorblockN{Shaolin Wang\IEEEauthorrefmark{1},
Yi Mei\IEEEauthorrefmark{2},
Haoyang Che\IEEEauthorrefmark{3},
He Jiang\IEEEauthorrefmark{4},
Shui Yu\IEEEauthorrefmark{1},
Ying Gu\IEEEauthorrefmark{1}}

\IEEEauthorblockA{\IEEEauthorrefmark{1}Geely Holding Group, China\\
Email: \{shaolin.wang1, Shui.Yu, ying.gu1\}@zeekrlife.com}

\IEEEauthorblockA{\IEEEauthorrefmark{2}Victoria University of Wellington, New Zealand\\
Email: yi.mei@ecs.vuw.ac.nz}

\IEEEauthorblockA{\IEEEauthorrefmark{3}Independent Researcher\\
Email: bigdatache@qq.com}

\IEEEauthorblockA{\IEEEauthorrefmark{4}Dalian University of Technology, China\\
Email: jianghe@dlut.edu.cn}
}

\begin{document}
\maketitle
\begin{abstract}
Software systems have traditionally been designed for human interaction, emphasizing graphical user interfaces, usability, and cognitive alignment with end users. However, recent advances in large language model (LLM)-based agents are changing the primary consumers of software systems. Increasingly, software is no longer only used by humans, but also invoked autonomously by AI agents through structured interfaces. In this paper, we argue that software engineering is undergoing a paradigm shift from \emph{human-oriented interfaces} to \emph{agent-oriented invocation systems}. We formalize the notion of \emph{agent interfaces}, introduce \emph{invocable capabilities} as the fundamental building blocks of AI-oriented software, and outline design principles for such systems, including machine interpretability, composability, and invocation reliability. We then discuss architectural and organizational implications of this shift, highlighting a transition from monolithic applications to capability-based systems that can be dynamically composed by AI agents. The paper aims to provide a conceptual foundation for the emerging paradigm of AI-native software design.
\end{abstract}


\section{Introduction}
Software systems have historically been designed with humans as their primary users. Interfaces are optimized for visual perception, usability, and interaction efficiency, typically through graphical user interfaces (GUIs). In such systems, humans interpret information, make decisions, and execute actions through predefined workflows.

This assumption is increasingly being challenged. The emergence of large language models (LLMs) and autonomous agents enables machines to interact with software systems directly. These agents can interpret natural language goals, plan multi-step tasks, invoke tools, and coordinate execution without continuous human intervention. As a result, software is no longer exclusively consumed through human-facing interfaces, but also through machine-driven interactions.

This transition introduces a fundamental mismatch. Existing systems are largely optimized for human interaction, making them difficult for AI agents to interpret, invoke, and compose reliably. Interfaces designed for human cognition often lack the structure, determinism, and explicit semantics required for machine usage. Recent work on tool-using language models, agent-computer interfaces, and agentic software engineering collectively points to a deeper shift in the role of software itself~\cite{schick2023toolformer,yang2024sweagent,hassan2025agenticse}.

In this paper, we argue that software design is undergoing a paradigm shift from \emph{human-oriented interfaces} to \emph{agent-oriented invocation systems}. Rather than treating AI as merely an assistant for writing software, we focus on a different question: how should software be designed when AI agents become first-class users of the system? To answer this question, we define the notion of an \emph{agent interface}, introduce \emph{invocable capabilities} as the fundamental abstraction of AI-oriented software, and propose a set of design principles for systems optimized for machine invocation.

The contributions of this paper are threefold:
\begin{itemize}[leftmargin=1.2em]
    \item We formalize the transition from human interfaces to agent interfaces as a software engineering paradigm shift.
    \item We introduce \emph{invocable capability} as the minimal abstraction for AI-oriented software functionality.
    \item We articulate design principles and architectural implications for capability-oriented software systems.
\end{itemize}

\section{Related Work}

Our argument is related to several active research directions on tool use, coding agents, and human--AI interaction, but it differs from them in its primary unit of analysis. Existing work mainly studies how AI systems can better use tools, operate computers, or assist software development. In contrast, we focus on how software systems themselves should be designed when AI agents become first-class users.

\subsection{Tool Use and Function Invocation in LLMs}

A foundational line of work studies how language models invoke external tools. Toolformer showed that language models can learn when to call tools, which tools to call, and what arguments to pass, using APIs such as calculators, search engines, and calendars~\cite{schick2023toolformer}. This line of research establishes that software tools can become part of a model's reasoning process.

However, these approaches primarily treat tools as augmentations to model capability. They assume that callable tools already exist and focus on improving how models use them. In contrast, our work asks a complementary question: \emph{how should software systems themselves be designed to support reliable and scalable machine invocation?}

\subsection{Agentic Programming and Software Engineering Agents}

A second line of work investigates autonomous agents for software engineering. SWE-agent demonstrates that carefully designed agent-computer interfaces (ACI) significantly improve an agent's ability to navigate repositories, edit code, and execute tests~\cite{yang2024sweagent}. More recent work frames this trend as agentic programming or agentic software engineering, emphasizing planning, tool integration, and execution monitoring in coding agents~\cite{wang2025agenticprogramming,roychoudhury2025agenticsoftware}.

These approaches share a common assumption: software systems are primarily designed for human interaction, and agents must be adapted to operate within them. In this sense, they focus on \emph{adapting agents to existing software environments}.

In contrast, we argue for a complementary but fundamentally different direction. Rather than adapting agents to human-oriented systems, we propose that software systems themselves should be redesigned to natively support machine invocation. Under this view, agent-computer interfaces can be interpreted as transitional mechanisms that expose the mismatch between human-centric design and agent-centric usage.

\subsection{Human--AI Interface and Collaboration Design}

Another related stream studies interfaces for human--AI collaboration. Prior work emphasizes properties such as transparency, controllability, and representational compatibility between humans and AI agents~\cite{demasi2026designproperties}. 

While this literature provides valuable insights into interaction design, it typically assumes that human interaction remains central. Our work considers a broader shift: when AI agents become primary users, the role of interfaces changes fundamentally, and software systems must be structured around machine invocation rather than human navigation.

\subsection{Intent-Centric and AI-Assisted Development}

Recent work on intent-centric development and "vibe coding" suggests that developers increasingly specify goals, curate context, and validate outcomes instead of directly implementing all logic~\cite{ge2025vibecodingsurvey}. This reflects a shift in how software is created.

Our contribution complements this perspective by shifting attention from software \emph{creation} to software \emph{structure and consumption}. We argue that when AI agents become first-class users, the fundamental unit of software shifts from user-facing features to agent-invocable capabilities.

\subsection{Summary}

Overall, prior work demonstrates that AI systems can use tools, interact with computers, and participate in software engineering workflows~\cite{schick2023toolformer,yang2024sweagent,wang2025agenticprogramming,roychoudhury2025agenticsoftware,ge2025vibecodingsurvey}. However, these approaches largely operate within the constraints of software designed for human users.

We argue that this perspective is incomplete. The emergence of AI agents as primary users calls for a rethinking of software design itself. We position \emph{agent interfaces} and \emph{invocable capabilities} as abstractions for this new paradigm, shifting the focus from adapting agents to systems toward designing systems for agents.

\section{Background and Motivation}
\subsection{Human-Oriented Software Systems}
Traditional software systems are designed around human interaction. Core design considerations include usability, visual clarity, and cognitive load. Interfaces such as web applications, mobile apps, and desktop GUIs enable users to navigate features, input data, and trigger actions. In this paradigm, interaction is perception-driven, functionality is organized into features and pages, and execution is human-initiated.

\subsection{Rise of AI Agents as Software Users}
Recent advances in LLMs enable agents to understand goals, reason over context, and autonomously invoke external tools. This capability is no longer limited to code generation or question answering. Agents increasingly operate over productivity applications, developer tools, search systems, and enterprise services. In effect, the ``user'' of software can now be an AI process that does not read screens the way humans do, but instead depends on structured interfaces, explicit contracts, and reliable side effects.

\subsection{The Core Mismatch}
Most contemporary software remains poorly suited for this mode of usage. Interfaces are often ambiguous, functional boundaries are coarse-grained, and execution semantics are hidden behind UI state or undocumented assumptions. These properties are tolerable for humans, who can infer missing context and recover from ambiguity, but they become major failure modes for autonomous agents.

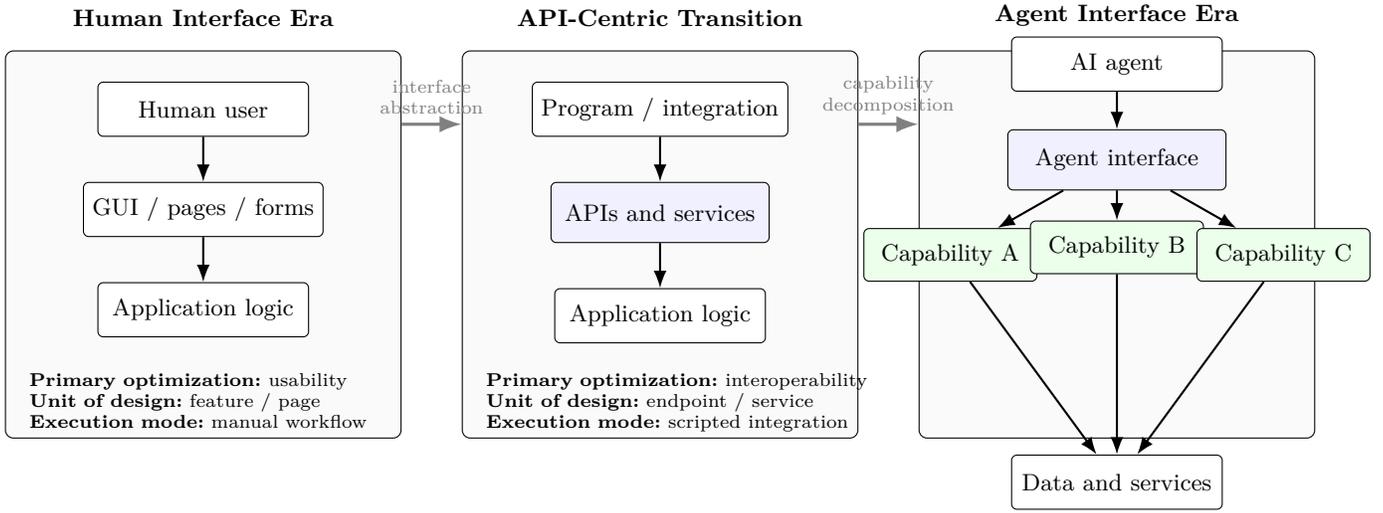
\begin{figure*}[t]
\centering
\begin{tikzpicture}[
    font=\small,
    >=Latex,
    node distance=7mm and 11mm,
    stage/.style={draw, rounded corners=3pt, minimum width=0.29\textwidth, minimum height=5.15cm, align=center, fill=gray!4},
    role/.style={draw, rounded corners=2pt, minimum width=2.8cm, minimum height=0.72cm, align=center, fill=white},
    api/.style={draw, rounded corners=2pt, minimum width=2.9cm, minimum height=0.8cm, align=center, fill=blue!6},
    capabilitybox/.style={draw, rounded corners=2pt, minimum width=2.3cm, minimum height=0.7cm, align=center, fill=green!8},
    label/.style={font=\bfseries\small},
    arrow/.style={->, thick}
]

% Stage containers
\node[stage] (s1) {};
\node[stage, right=8mm of s1] (s2) {};
\node[stage, right=8mm of s2] (s3) {};

% Stage titles
\node[label, above=2mm of s1.north] {Human Interface Era};
\node[label, above=2mm of s2.north] {API-Centric Transition};
\node[label, above=2mm of s3.north] {Agent Interface Era};

% Stage 1
\node[role] (u1) at ([yshift=18mm]s1.center) {Human user};
\node[role, below=6mm of u1] (ui1) {GUI / pages / forms};
\node[role, below=6mm of ui1] (sys1) {Application logic};
\draw[arrow] (u1) -- (ui1);
\draw[arrow] (ui1) -- (sys1);
\node[align=left, anchor=north west, font=\scriptsize, text width=7.4cm] at ([xshift=2mm,yshift=10mm]s1.south west) {\textbf{Primary optimization:} usability\\\textbf{Unit of design:} feature / page\\\textbf{Execution mode:} manual workflow};

% Stage 2
\node[role] (p2) at ([yshift=18mm]s2.center) {Program / integration};
\node[api, below=6mm of p2] (api2) {APIs and services};
\node[role, below=6mm of api2] (sys2) {Application logic};
\draw[arrow] (p2) -- (api2);
\draw[arrow] (api2) -- (sys2);
\node[align=left, anchor=north west, font=\scriptsize, text width=7.4cm] at ([xshift=2mm,yshift=10mm]s2.south west) {\textbf{Primary optimization:} interoperability\\\textbf{Unit of design:} endpoint / service\\\textbf{Execution mode:} scripted integration};

% Stage 3
\node[role] (a3) at ([yshift=24mm]s3.center) {AI agent};
\node[api, below=5mm of a3] (ai3) {Agent interface};
\node[capabilitybox, below left=5mm and -4mm of ai3] (c31) {Capability A};
\node[capabilitybox, below=4mm of ai3] (c32) {Capability B};
\node[capabilitybox, below right=5mm and -4mm of ai3] (c33) {Capability C};
\node[role, below=24mm of c32] (db3) {Data and services};
\draw[arrow] (a3) -- (ai3);
\draw[arrow] (ai3) -- (c31);
\draw[arrow] (ai3) -- (c32);
\draw[arrow] (ai3) -- (c33);
\draw[arrow] (c31) -- (db3);
\draw[arrow] (c32) -- (db3);
\draw[arrow] (c33) -- (db3);

% Flow arrows between stages
\draw[->, very thick, gray] ([yshift=16mm]s1.east) -- node[above, font=\scriptsize, align=center] {interface\\abstraction} ([yshift=16mm]s2.west);
\draw[->, very thick, gray] ([yshift=16mm]s2.east) -- node[above, font=\scriptsize, align=center] {capability\\decomposition} ([yshift=16mm]s3.west);

\end{tikzpicture}
\caption{Evolution of software interaction paradigms. The proposed shift is not merely from GUI to API, but from feature-centric applications consumed by humans to capability-oriented systems invoked and dynamically composed by AI agents.}
\label{fig:paradigm_shift}
\end{figure*}

\section{From Human Interfaces to Agent Interfaces}
We define an \emph{agent interface} as a software interaction layer designed for machine invocation, characterized by structured inputs, explicit semantics, and deterministic execution. Unlike a human interface, which is optimized for navigation and usability, an agent interface is optimized for interpretability, reliable invocation, and composability.

Figure~\ref{fig:paradigm_shift} summarizes the proposed paradigm shift. Human-oriented systems typically expose functionality through pages, forms, and workflows. API-centric systems partially decouple functionality from the UI, but still often reflect service boundaries chosen for human-led integration. Agent-oriented systems go further: they expose software as a collection of invocable capabilities that can be selected and composed dynamically by an AI agent.

Table~\ref{tab:comparison} compares the two paradigms.

\begin{table}[h]
\centering
\caption{Human interfaces versus agent interfaces}
\label{tab:comparison}
\renewcommand{\arraystretch}{1.15}
\begin{tabular}{p{2.2cm}p{2.7cm}p{2.7cm}}
\toprule
\textbf{Dimension} & \textbf{Human Interface} & \textbf{Agent Interface} \\
\midrule
Interaction mode & GUI-based navigation & API / tool invocation \\
Input style & Ambiguous, flexible & Structured, validated \\
Execution & Manual & Autonomous \\
Optimization target & Usability & Reliability \\
Composition & Predefined workflow & Dynamic orchestration \\
Primary abstraction & Feature / page & Capability / contract \\
\bottomrule
\end{tabular}
\end{table}

The shift can therefore be stated more precisely: software is moving from \emph{perception-driven interaction} to \emph{invocation-driven execution}. This is not only an interface change, but also a change in software architecture, modularization strategy, and evaluation criteria.

\section{Invocable Capabilities as Fundamental Units}
To make agent interfaces concrete, we introduce the concept of an \emph{invocable capability}.

\textbf{Definition.} An invocable capability is a minimal, self-contained unit of functionality that exposes a structured interface and can be reliably executed by an AI agent without human intervention.

This abstraction differs from a user-facing feature in several important ways.

First, the input and output of an invocable capability must be explicitly structured. A capability should not rely on hidden UI state, informal conventions, or tacit assumptions. Second, the capability should be self-contained: it can be executed independently once its inputs are provided. Third, it should exhibit a high degree of predictability. Given the same inputs and environment, repeated invocations should produce behavior that is sufficiently consistent for planning and orchestration.

These properties make invocable capabilities reusable across tasks and environments. Instead of binding software functionality to specific screens or rigid workflows, systems can expose fine-grained units that AI agents select and combine as needed. As such, the fundamental unit of software shifts from \emph{feature} to \emph{capability}.

\section{Design Principles for AI-Oriented Software}
The proposed paradigm suggests several design principles for software intended to support AI agents as first-class users.

\subsection{Machine Interpretability}
Software should expose functionality in a form that agents can reliably interpret. This includes explicit operation names, typed parameters, semantic descriptions, and clear state transitions. Ambiguity that is tolerable in human interfaces often becomes a source of failure for autonomous agents.

\subsection{Composable Capability Design}
Systems should prefer modular, reusable capabilities over coarse-grained features. A composable capability can be combined with other capabilities to support flexible problem solving. This is essential when agents dynamically plan execution paths rather than follow a fixed navigation flow.

\subsection{Explicit Interface Contracts}
All interactions should be governed by explicit contracts. Contracts specify input schemas, output schemas, constraints, validation rules, side effects, and failure modes. For human users, many of these details can remain implicit; for agents, they should be part of the interface definition.

\subsection{Invocation Reliability}
Autonomous invocation requires predictable and robust behavior. A capability should minimize hidden dependencies, provide transparent error handling, and support safe retries whenever possible. Reliability therefore becomes a first-class design objective, not merely an implementation quality attribute.

\subsection{Context Compatibility}
Because many AI agents operate within limited context windows, interfaces should be concise and semantically economical. Capability descriptions should be sufficiently rich to support correct use while remaining compact enough for efficient inclusion in agent context.

\section{Discussion and Future Directions}
The proposed perspective has several implications for software engineering practice.

\subsection{From Monolithic Applications to Capability-Based Systems}
We envision a future in which software is no longer organized primarily as monolithic applications with layered human workflows. Instead, systems will increasingly expose collections of atomic, invocable capabilities that AI agents can discover, select, and combine at runtime. UI layers may remain important, but they will often become secondary interfaces for supervision, debugging, and exception handling.

\subsection{Changes in Architecture and Product Form}
This transition encourages a move from feature-centric product design to capability-centric platform design. In such systems, value is not determined solely by what a human can click through, but also by what an agent can reliably invoke. Discoverability, orchestration, capability granularity, and interface standardization become central architectural concerns.

\subsection{Open Challenges}
Several challenges remain open. First, capability granularity is non-trivial: overly coarse capabilities limit flexibility, while overly fine capabilities may create planning overhead. Second, standardization of agent interfaces remains immature. Third, autonomous invocation introduces new concerns related to security, misuse, and policy enforcement. Finally, many real systems must serve both human and AI users, creating a need for dual-audience design rather than a complete replacement of one interface type with another.

\section{Conclusion}
The emergence of AI agents as active software users introduces a fundamental shift in software engineering. Systems are no longer designed solely for human interaction, but increasingly for machine invocation. In response, we argued for a transition from human interfaces to agent interfaces, introduced invocable capabilities as the fundamental abstraction for AI-oriented software, and outlined design principles and architectural implications for this emerging paradigm.

Our claim is not that graphical interfaces will disappear, nor that APIs alone solve the problem. Rather, the key shift is that software must increasingly be structured so that autonomous agents can understand, invoke, and compose functionality reliably. We believe this change will redefine how software is designed, modularized, and evaluated, marking an important step toward AI-native software systems.

\balance
\bibliographystyle{IEEEtran}
\bibliography{references}

\end{document}